# The Future of Time: UTC and the Leap Second

*Earth's clocks have always provided Sun time. But will that continue?*

David Finkleman, Steve Allen, John H. Seago, Rob Seaman and P. Kenneth Seidelmann[*]

[†]*"From ages immemorial, the means universally used to mark off the passage of time has been the apparent motions of the celestial bodies; and the measurement of time whether by hours of day and night, or the longer intervals involved in calendrical and chronological reckoning, is still dependent upon the celestial motions. Other means, such as clocks, are only auxiliaries or intermediaries."*

American astronomers Edgar Woolard and Gerald Clemence published those words in 1966, not that long ago in the history of human timekeeping. But what was certain then—the necessity of linking civil time to the motion of celestial bodies— may soon be abandoned. In January 2012, a United Nations–affiliated organization known as the Radiocommunication Sector of the International Telecommunications Union (ITU-R) could permanently break this link by redefining Coordinated Universal Time.

Coordinated Universal Time is better known by its international acronym UTC, the modern successor to traditional Greenwich Mean Time. It is the basis for legal timekeeping in most of the world, including the United States. It is maintained by a continuous count of *SI seconds*, the metric system's fundamental unit of duration. By definition, a SI second is the duration of 9,192,631,770 periods of radiation from the transition between two hyperfine levels of the ground state of the cesium-133 atom, a physical phenomenon distinct from the rotation of Earth.

Before atomic timekeeping, clocks were set to the skies. But starting in 1972, radio signals began broadcasting atomic seconds. Since then, *leap seconds* have occasionally been added to that stream of atomic seconds to keep the signals synchronized with the actual rotation of Earth. Adjustments are needed because Earth's rotation is slightly less regular and a bit slower on average than cesium-133's quantum-scale rhythms.

Leap seconds have remained a means to an end—that is, to reconcile and combine two entirely different yet useful notions of timekeeping. But the ITU-R proposal would cease issuing leap seconds entirely. Clocks everywhere—on your wall, wrist, phone and computer—would begin to diverge from the heavens. The change would bring scientific, technological, legal, philosophical and social implications too. These range from abandoning the requirement to preserve knowledge of Earth's rotation through timekeeping, to whether the word *day* will mean "one turn of the Earth" versus "794,243,384,928,000 cycles of cesium-133 radiation." As scientists and engineers engaged in astronautics, space navigation and astronomy, we seek to draw attention to this unprecedented situation.

[*] David Finkleman is the senior scientist at the Center for Space Standards and Innovation and convenor of the International Standards Organization's Space Operations Working Group. Steve Allen is a programmer-analyst at the University of California's Lick Observatory. John Seago is an astrodynamics engineer at Analytical Graphics, Inc. Rob Seaman is a senior software systems engineer at the National Optical Astronomy Observatory. Ken Seidelmann is a research professor of astronomy at the University of Virginia and the former director of astrometry at the U.S. Naval Observatory. Address for Finkleman: 7150 Campus Drive, Colorado Springs, CO 80920. E-mail: dfinkleman@centerforspace.com

**Calendars: Counting the Days**

A calendar is a system for labeling the sequence of days over long periods in a manner consistent with lunar and solar cycles. The average duration of a solar day, determined by Earth's rotation, does not divide neatly into the orbital periods for the Sun and the Moon. The seasons would shift a little each year if calendars were not adjusted regularly. The ancients knew that some calendar years must be leap years. Many ancient cultures were able to deduce—even with crude measuring devices—that 365¼ Earth rotations equal one Earth revolution about the Sun. Thus the calendar established by Julius Caesar in 45 B.C. had one leap day every four years.

We now know that 365.25 days per year is an insufficient approximation, long by approximately 0.002 percent on average. The discrepancy had grown to 10 days by the time Pope Gregory XIII tackled the problem 16 centuries later. To compensate, Friday, October 15, 1582 was decreed to directly follow Thursday, October 4. To adjust the calendar to the date of the equinox and Easter, the new Gregorian calendar scheme omitted three Julian leap days every four centuries. Different countries and religions adopted the reforms at different times and in different ways. In 1712, Sweden added a February 30 to its national calendar. Great Britain and its colonies resisted Gregorian reforms for two centuries. For example, George Washington was born on February 11, 1731 O.S. (Old Style). The United States celebrates his birthday each year on February 22 because an 11-day deficit was corrected during his lifetime. The date of Washington's Birthday on the calendar may be arguable, but the astronomical timing of his birth is not.

The Gregorian calendar is used internationally for official purposes. Some national traditions and religions use a lunar calendar averaging 354 days (twelve lunations) per calendar year, which allows the start of the calendar year to drift into different seasons. Other lunar calendar systems maintain proximity to the seasons by introducing an occasional "leap month"—thus some years have thirteen lunations. Cultures that hold to religious, lunar or luni-solar calendars for civil purposes require concordance between their calendars and the Gregorian calendar for commerce. The Chinese calendar is luni-solar, combining cycles of both the Moon and the Sun; a table is needed to map to the corresponding moveable dates relative to a purely solar calendar.

Correctly interpreting any historical date requires understanding of both today's calendar and the calendar used at the time and place of the event. To program such calendar conversions into a computer requires a tabular database to collect the rules and information for different places and times, including special cases such as Daylight Saving Time or Summer Time transitions. Intercalary adjustments are inevitable, and accommodations for them have been and must continue to be maintained throughout history.

The leap second is analogous but occurs on a much more precise scale. Modern measurement and technical capabilities make it necessary to accommodate extremely small discrepancies between the time it takes Earth to rotate and atomic-clock time. In fact, the extreme precision of atomic frequency standards makes these discrepancies more apparent. Just as a database of special cases is required for calendars, a table of leap seconds is required for clocks.

**Clocks: Down to Seconds**

Whereas calendars count the days, clocks divide them. The regulated flows of sand and water, swinging pendulums, vibrating crystals and atomic radiation have all been used as a basis for various types of clocks, carving the time of day into shorter and more uniform increments. The hour originated in ancient Egypt, which designated 12-hour days and 12-hour nights, even though the lengths of each vary throughout the year. We inherited the division of hours into units of 60—an easily fractionalized number—from sexagesimal Babylonian fractions. The oldest and most widespread understanding of a second is still a sixtieth of minute, with 60 minutes in an hour, and 24 hours in a day. This is the traditional meaning of the word "second"—that is, the fraction 1/86,400 of a day (86,400 = 24×60×60).

There are several kinds of astronomical days, and multiple terms exist to describe them. While Earth orbits the Sun once per year, it also spins on its axis relative to the "fixed" stars. The *apparent solar day*, the period between when the Sun is on the meridian and when it next returns there, is kept by a simple sundial. It is not uniform, and varies by approximately 30 minutes[*] in the course of a year. A *sidereal day* is one Earth rotation relative to the equinox and stars (*see Figure 2*) and is more uniform because it is not distorted by Earth's orbit around the Sun.

Given a measure of sidereal days per year, it is straightforward to determine that the apparent solar day is approximately four minutes longer than the sidereal day on average. This leads to the definition of *mean solar time*, which is the solar time that is best suited for use with clocks and that best tracks Earth's rotation relative to the Sun *on average*. On a day-to-day basis, one may notice variations in all this astronomical motion, which tends to average out over the course of the year. This is best exemplified in the *analemma*—the path across the sky that the Sun appears to make if viewed daily at the same time over the course of the year. Its figure-eight-like pattern is due to the eccentricity of Earth's orbit and the inclination of Earth's axis relative to the ecliptic—that is, to the mean orbital plane of Earth.

In the late 19th century, Simon Newcomb was director of the U.S. Nautical Almanac Office, which published astronomical data needed by astronomers, surveyors and navigators. He calibrated the duration of the mean solar day to unprecedented accuracy by using observations of the Sun and planets recorded back to 1750. Mean solar time was the independent variable in the equations used to calculate the tables of relative motion for the Sun and planets. Newcomb suspected that Earth's rotation was slightly irregular because his tabular lunar positions were slightly unpredictable using mean solar time. The theory of tides also predicted that the Moon should rob a little angular momentum from Earth each century, gradually slowing Earth down and enlarging the Moon's orbit slightly. Newcomb averaged more than 150 years of such variation to achieve his equation, which was adopted as an international standard.

By the 1930s, astronomers had confirmed that the independent variable designated as time by Newcomb's solar-system theory did not correlate precisely with time as observed by Earth's (slightly irregular) rotation. These two concepts were separated for some precision tasks; the direct astronomical measure of Earth's rotation that was once known as "mean solar time at Greenwich" became known as Universal Time (UT), whereas Newcomb's independent variable eventually became known as Ephemeris Time. At any future epoch, the difference between the observed location of the Moon or a planet and the location predicted by Newcomb's tables could be used to infer the difference between Ephemeris Time and Universal Time. Gerald Clemence, based at the U.S. Nautical Almanac Office decades after Newcomb, in 1948 clarified that the purpose of Ephemeris Time was "for the convenience of astronomers and other scientists only" as it was "logical to continue the use of mean solar time [Universal Time] for civil purposes." By 1960, the theoretically uniform *ephemeris second* had been adopted as the fundamental unit of duration within the International System of Units (known internationally as the *Système International d'Unités*, or SI). The constancy of this SI unit was best suited for tasks such as solar-system theories and the calibration of electromagnetic frequencies. But the precise determination of Ephemeris Time relative to Universal Time from celestial observations was slow and cumbersome. Already, atomic resonators were capable of ultraprecise measures of time intervals. In 1958, Louis Essen of the National Physical Laboratory in England and William Markowitz of the Time Service at the U.S. Naval Observatory determined the frequency of radiation emissions from hyperfine transitions of cesium-133 atoms that best matched the duration of the existing ephemeris second. By 1968, the definition of the SI second officially changed from a unit of duration based on Ephemeris Time to one based on atomic frequency. Within a few years, the sequence of atomic seconds maintained and coordinated by the International Bureau of Weights and Measures became its own laboratory timescale, known as the *Temps Atomique International* (TAI).

---

[*] *Addendum*: Apparent time-of-day is described by the *equation of time*, which is accumulated day-by-day from small differences between the apparent length-of-day and the length of the mean solar day. Length-of-day varies by only about plus or minus a half-minute through the course of a year.

**UTC and Leap Seconds**

Many scientific, engineering and technical applications rely on the constant duration of atomic SI seconds. But there have never been exactly 86,400 SI seconds in one solar day. The difference is only milliseconds per day, but the counting of atomic seconds leads to a noticeably different time compared to mean solar seconds of Universal Time. Internationally, many laws, regulations, public expectations and technologies require that civil timekeeping follow Universal Time. Celestial navigators need a source of Universal Time accurate to less than 0.2 seconds to fix their positions on Earth. Sighting a star or a satellite from Earth requires clocks that accurately track Earth's orientation via UT.

Throughout the 1960s, radio signals were able to broadcast Universal Time by introducing time steps of less than 0.1 seconds and by varying the length of the broadcast second compared to stable atomic-frequency standards. But this practice made it difficult to recover precision time intervals or calibrate electromagnetic frequencies from these broadcasts, because the apparent length of a "broadcast second" varied with respect to fundamental frequency standards.

In 1972, leap seconds were introduced into UTC. The concept of UTC was refined by the commissions of the International Astronomical Union (IAU) during IAU General Assembly meetings in 1970 and 1973. The official normative description of UTC was implemented through the Consultative Committee on International Radio (CCIR), a forerunner of the ITU-R. Ever since, UTC has only mandated occasional adjustments. Integer leap seconds sustain the relationship between atomic seconds and the precise version of Universal Time known as UT1. UT1 best represents the instantaneous orientation of Earth.

Leap seconds are experienced at the same moment worldwide. Because of time zones, a leap second occurs an hour earlier in local time for each time zone stepped to the west and an hour later for each time zone to the east of Greenwich, England. To date, 24 leap seconds have been introduced. Today, UTC can be obtained to various levels of fidelity through many different means, including radio signals, the U.S. Global Positioning System (GPS), communication- and weather-satellite broadcasts, and Internet-timing protocols. However, when UTC was instituted, shortwave radio broadcasts were the primary means for distributing timing signals in real time. Because of this, the ITU-R became responsible for maintaining UTC's transmission guidelines, even though ITU-R had no involvement with timescale development. The International Telecommunications Union is chartered under the United Nations as an administrative regulatory body, meaning that it has neither the authority nor the power to ensure that the Coordinated Universal Time scale is generated, distributed or employed properly. In fact, no single organization governs this. But ITU findings are highly influential with policymaking government bodies worldwide. The sequence of atomic seconds is actually maintained by the International Bureau of Weights and Measures. The International Earth Rotation and Reference Systems Service determines the need for a leap second while monitoring Earth's rotation with exquisite precision relative to distant cosmic radio sources. It schedules a leap second when the difference between UT1 and UTC approaches or exceeds −0.5 seconds, usually six months in advance and always as the last second of the last day of a month (so far only June or December). A positive leap second is labeled as 23h 59m 60s, according to convention, which increases the duration of one UTC day while sustaining continuous, unambiguous time marking.

**At the Crossroads**

Most individuals worldwide are unaware of UTC and its origins and probably assume that a calendar day equals exactly 24×60×60 seconds, as is ingrained in us from childhood. Even software designers and hardware manufacturers are not always aware of UTC's prescriptions for leap seconds. Sometimes leap seconds are known about but simply ignored. Typically this lapse is inconsequential by design, as there are limited applications that truly require time stamps to be determined relative to UTC or require zone time to be known to the second. In particular, computers are not accurate clocks. Their internal oscillators can drift by minutes per day. Most computers are programmed to regularly synchronize with an accurate time source. Many networked devices simply forgo any leap-second insertion and allow incorrect time labeling until the system clock has been reset against a timing service. Leap seconds might not matter to such applications, but they do not impair them.

Leap seconds can create problems in systems that require precise synchronization but are not designed to handle minutes with an extra second. Those who share data among distributed computer systems are concerned that the precise data sequencing might be confused if some elements of the network are not aware of the leap second. Part of the problem is the notification system in use. Leap second announcements are officially published as text in IERS Bulletin C, shared much like 19th-century diplomatic communications. Someone must then convert the information into a form that computers can understand. Some systems such as GPS provide electronic notice of upcoming leap seconds shortly after their announcement, about six months in advance. A GPS receiver may be the most reliable and suitable method for automatically discovering when leap seconds will occur, although time kept by the GPS system includes no leap seconds and is now 15 seconds ahead of UTC. The U.S. National Institute of Standards and Technology's shortwave broadcasts provide leap-second notifications up to 30 days in advance. Some protocols for networked devices allow for only about a one-minute warning, too little notice for some purposes. At sites where updates are not automated, notice may arrive too early or too late.

Analog clocks cannot accommodate a 61-second minute, but, in principle, digital clocks and software-based clocks can. Displaying a negative leap second (the second following 23:59:58 would be 00:00:00 the next day) should be even easier, although one has not yet occurred. A lack of inexpensive hardware and clock circuitry to correctly accommodate an extra second— such as 23h 59m 60s—has resulted in imaginatively unconventional ways to represent UTC or zone time near a leap second, which can also cause problems in synchronizing computer clocks near a leap second.

When UTC with leap seconds was first introduced, it offered tremendous convenience and advantage to the telecommunication industry. A generation later, its leap seconds were much less appreciated. In 2001, the ITU-R elected to study the question of leap seconds, and a special ITU-R study group recommended their cessation while retaining the name "Coordinated Universal Time" and the abbreviation "UTC."

In most countries, including the United States, UTC remains the statutory basis for civil timekeeping, and it is a convention that should not be changed capriciously. Statutes that address timekeeping did not always anticipate material changes in UTC, nor did they define it with great detail. ITU-R's Radiocommunication Assembly, meeting in Geneva in January 2012, will vote on the recommendation. If it is approved, leap seconds will be discontinued in 2018, and Coordinated Universal Time will no longer be coordinated with Universal Time.

**Is the Sky Falling?**

No calendar system endures indefinitely; history has taught us that every system of timekeeping has a limited lifespan. Even the Gregorian calendar will at some point require that we make more leap-day adjustments or switch to another method, because its rules are still just approximations of real astronomical relations.

Similarly, UTC will not function a thousand years from now because more leap seconds will be needed per annum than there are months in which to add them. This is because the length of the day is expected to increase over the millennia. Although in recent years the day has been getting shorter on average, over the very long term the length of the astronomical day is expected to grow a few milliseconds more with each passing century. This is a consequence of lunar tides tugging against the spinning Earth. We have leap seconds because the length of the day is about 86,400.001 SI seconds; thus we need roughly one leap second every two or three years (1 millisecond per day × 1000 days). In about 50,000 years, the astronomical day is expected to be about 86,401 SI seconds. A leap second then would be required daily, except that timekeeping should change materially before that.

It is interesting to speculate how posterity might adapt to this situation. Clocks say as much about humankind as any technology we have. But timekeeping is an issue for the here and now, and it will continue to be critical to technological and social infrastructure through all the centuries to come. We will always need to distinguish night from day, to divide our working days hour by hour, to time our most treasured events. But if we formally sever timekeeping from the motion of the sky, it is unclear how we could ever rejoin the two. The cessation of leap seconds would suddenly remove the requirement that timekeeping and telecommunications equipment have built-in capacities to maintain intercalary adjustments. That would introduce technological roadblocks for realigning our global timekeeping practices with the heavens.

The chief advantage for repealing leap seconds today is that the logistical overhead needed to keep track of and display them will no longer be required by systems that do not need astronomical time to function. This would provide a notable simplification to telecommunications and some electronic navigation systems, but those who deal with historically time-stamped information will be handicapped. Reprogramming of operational software that already presumes UT1 and UTC are always within a second of each other would be required, and some space operations and astronomical applications would need to distinguish between the UTC without leap seconds and UT1. There would also be a need for statutory and regulatory changes to national legal systems for which astronomical time is the (explicit or implicit) standard. And a review, update and republication of documents and standards discussing timekeeping and UTC would be necessary. Because the difference between apparent time of day and clock time would grow without leap seconds, a longer-duration adjustment would be necessary in the future. That would be less easy to overlook compared to today's leap seconds. Hardware would still need to represent such an adjustment in an untraditional way. Without *any* clock adjustments, regional governments in a distant time would eventually be compelled to redefine the time zones, and the International Date Line would have to be moved to avoid having the calendar date change during normal daylight hours.

There is no consensus in industry, academia or among governments regarding the next best step for civil timekeeping, Coordinated Universal Time and leap seconds. The confounding aspects of UTC, when they exist, are more often seen during its implementation than in establishing its definition. And even if leap seconds ceased, the underlying issue of how to reconcile the steady advance of SI seconds with the astronomical day will never go away. Will deliberations at the ITU-R Radiocommunication Assembly in January 2012 resolve or cloud these issues?

Only time will tell.

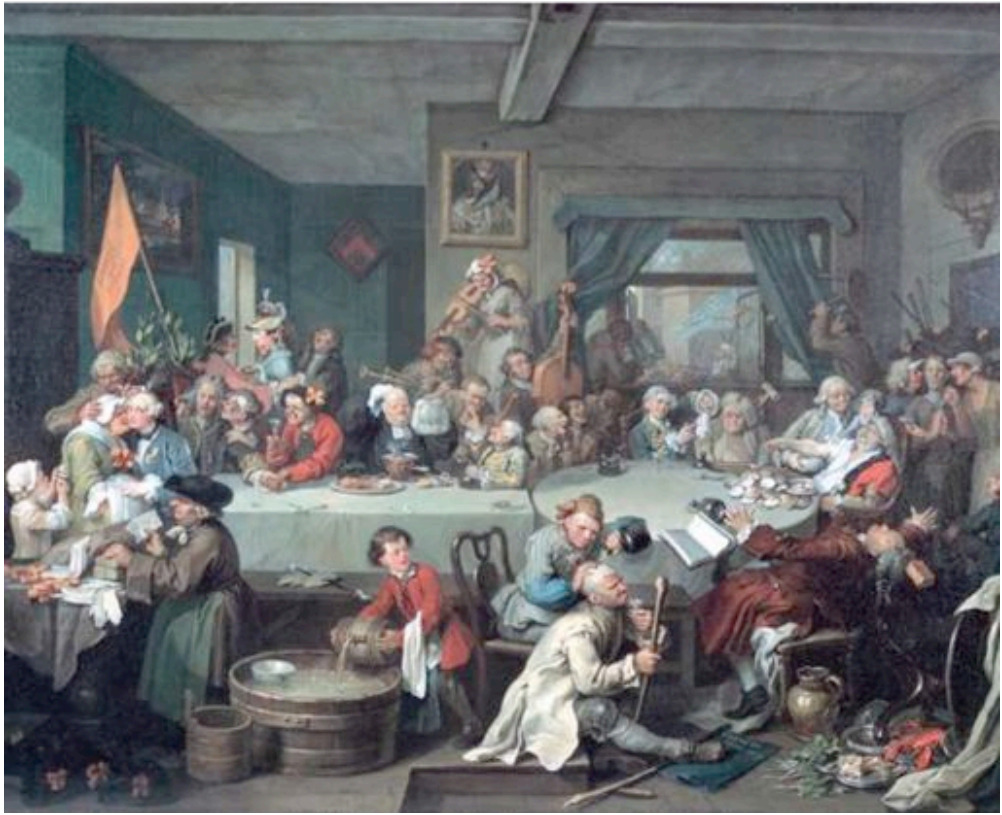

Figure 1. The 1754 painting *An Election Entertainment*<sup>*</sup> by the satirist William Hogarth depicts a fictional Whig election banquet. Under the boot of one participant, a banner (*magnified below*) reads: "Give Us Our Eleven Days." The demand refers to unrest in England over advancing the calendar 11 days in 1752 to adopt the Gregorian calendar. People have tinkered with timekeeping techniques for millennia to keep concordance with the heavens. This article explores a proposal in circulation today to eliminate leap seconds, which are adjustments to atomic-clock timekeeping that keep civil time synchronized with Earth's rotation. (Image courtesy of the Trustees of Sir John Soane's Museum.)

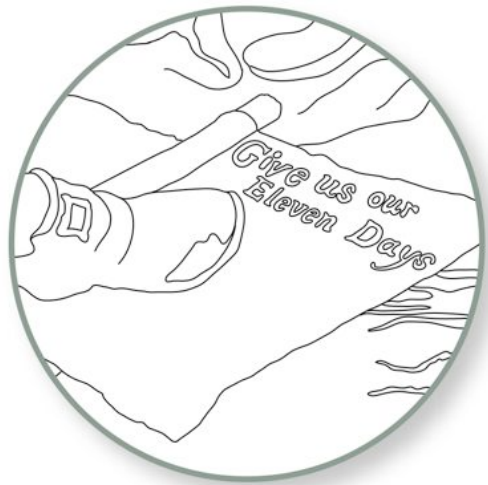

---

* The image of the 1754 painting *An Election Entertainment* by William Hogarth is copyrighted by Sir John Soane's Museum, London.

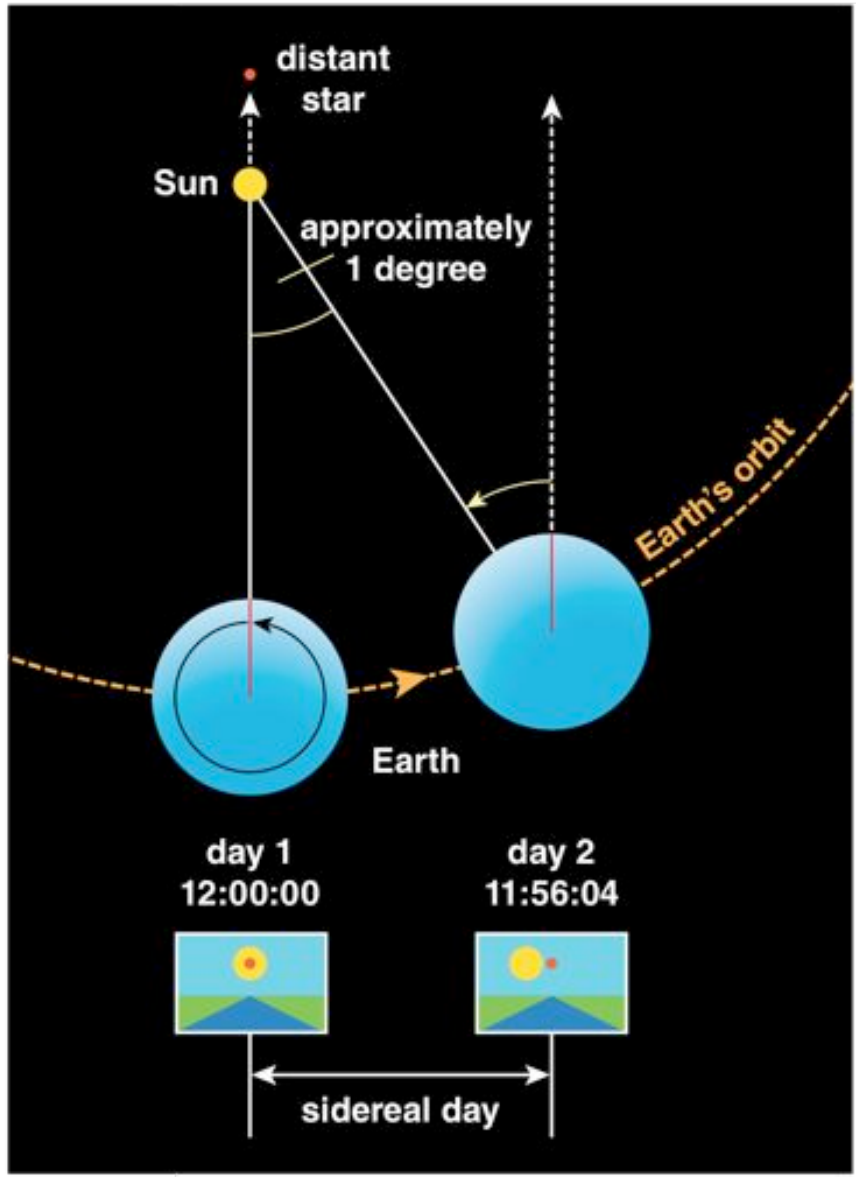

Figure 2. When measuring the length of one Earth rotation, apparent solar time is variable since Earth's orbit around the Sun affects how long it takes for the Sun to reappear above any spot on Earth. Sidereal time is more uniform because it takes advantage of the fact that a more distant star or other object will have less relative motion and thus appear stationary. The mean solar day is the sidereal day adjusted for Earth's annual orbit around the Sun.

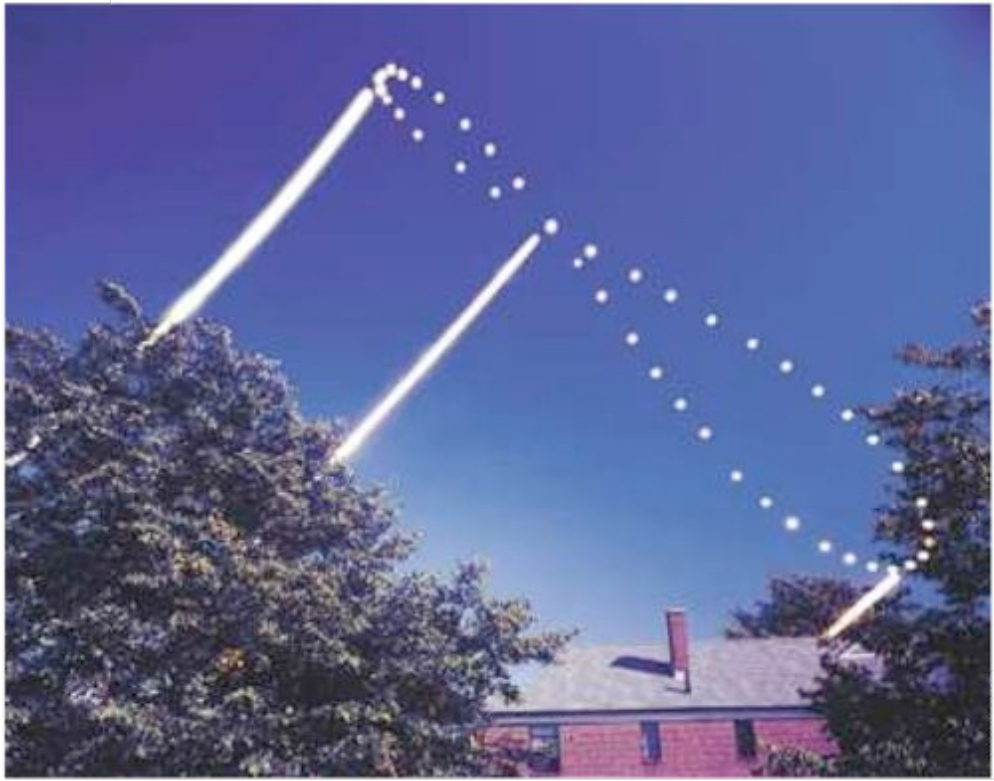

Figure 3. An analemma is an example of the variation in the Sun's appearance in our sky. It is the shape that the sun appears to make over a year when viewed from the same spot at the same time of day. Earth's elliptical orbit and the tilt of its axis produce the shifting view. Dennis di Cicco of *Sky & Telescope* magazine made this copyrighted photograph, the first ever of the analemma. Between February 1978 and February 1979, he made 45 exposures on a single piece of film loaded into a camera mounted in an east-facing window of his home in Watertown, Massachusetts. He made 44 exposures of the sun and one more to fill the rest of the frame with the foreground scene when the Sun was in the western sky and out of the camera's field of view. The streaks are time exposures of the sun's path from the horizon at the summer (*left*) and winter (*right*) solstices, and at the points (*center*) where the paths intersected.

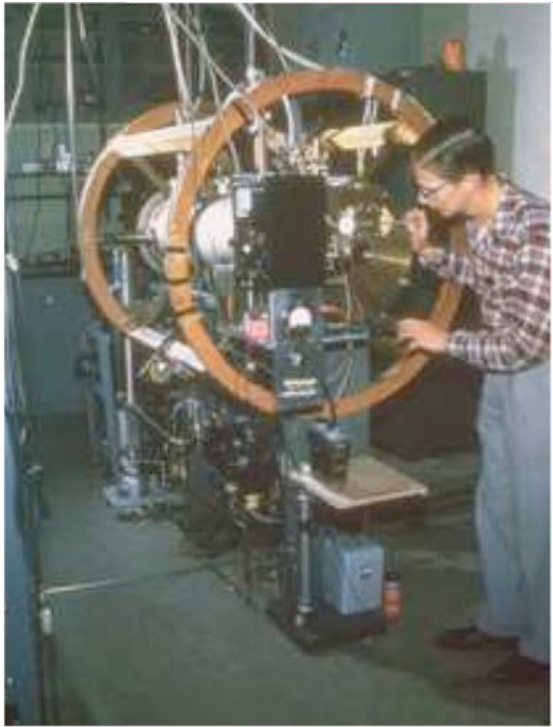
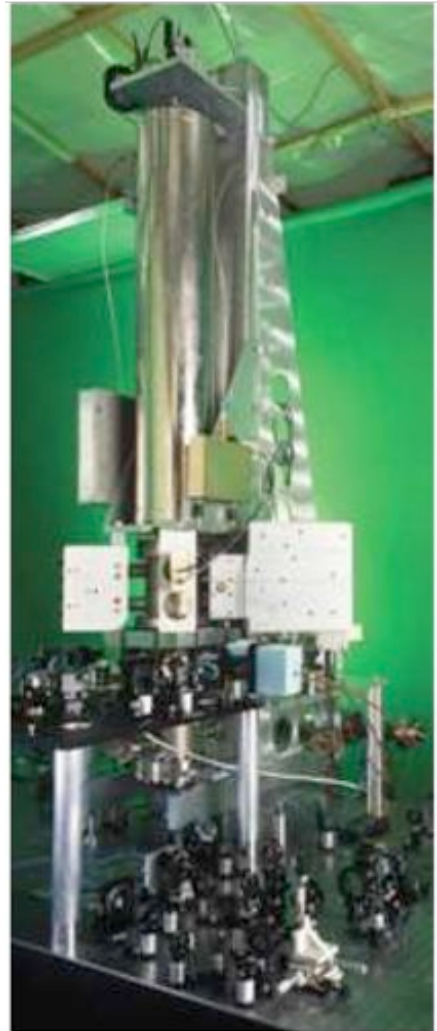

Figure 4. The U.S. National Institute of Standards and Technology (NIST) achieved the first accurate measurement of the frequency of cesium resonance in 1952. It used the instrument NBS-1, pictured above. At right is NIST-F1, the United States' present-day civilian time and frequency standard. It is believed to be accurate in measuring SI seconds at an error rate of only one second every 100 million years. (Photographs courtesy of NIST. Image at right by Geoffrey Wheeler.)

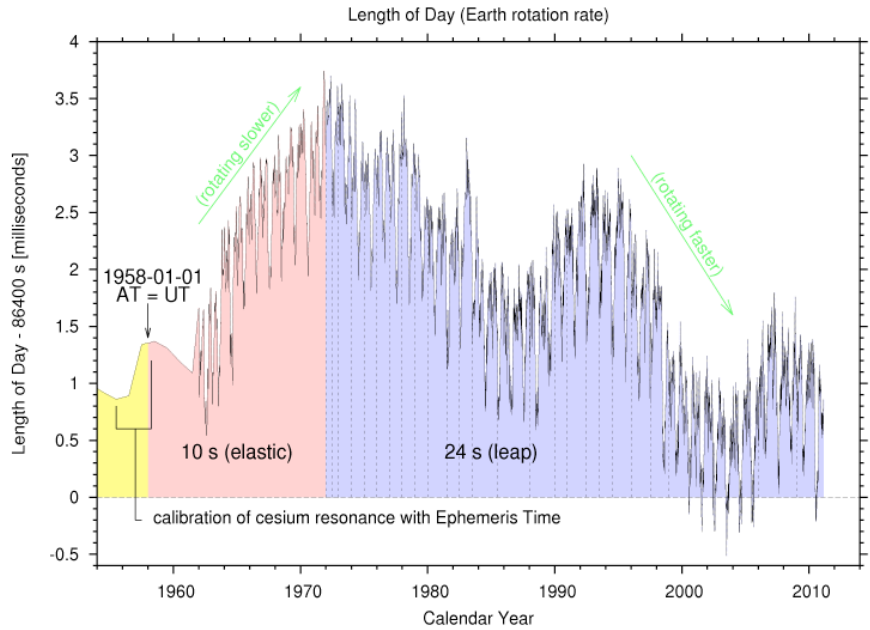

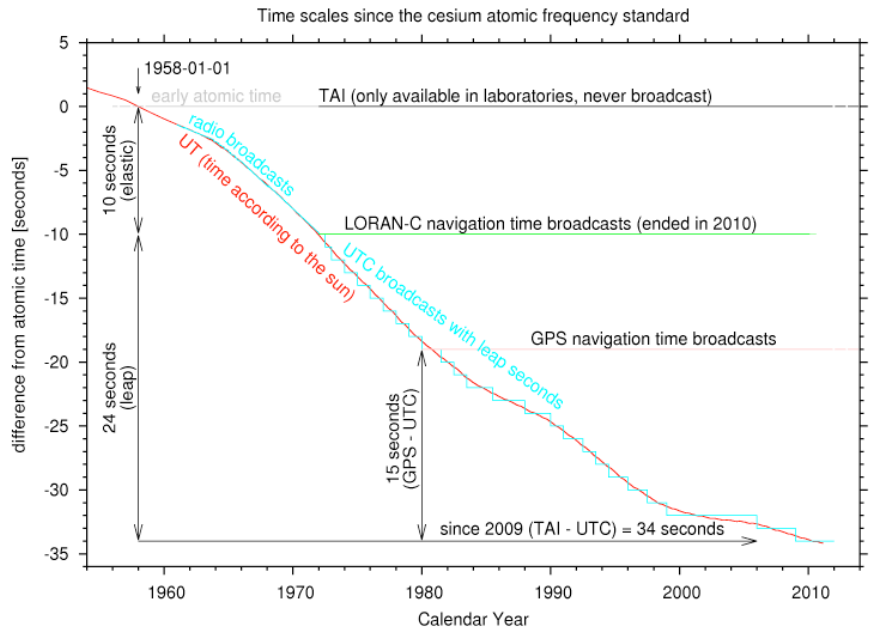

Figure 5. Earth's rotation is not as consistent as the resonance of cesium atoms. The very slight variation over time is captured in the graph at top, which shows the difference between 86,400 SI seconds and the actual number of seconds it took the planet to turn on average over decades. While Earth's rotation slows over centuries, slight variations occur over decades. The lower graph shows the transition from Universal Time to Coordinated Universal Time in 1972, along with the 24 leap seconds required since then to synchronize atomic time with Earth's actual rotations. GPS timekeeping includes no leap seconds and is now 15 seconds ahead of UTC. (Graphs adapted from figures produced by co-author Steve Allen.)

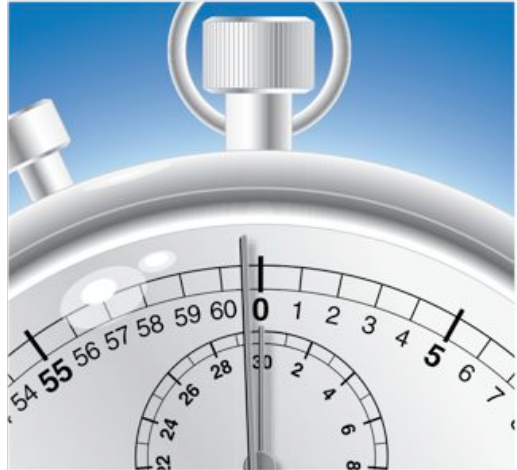

Figure 6. Incorporating leap seconds, which are scheduled rarely, has not always gone smoothly. It is not possible to accommodate them using a standard analog clock, for instance, although such precision is not as important in those timepieces. But not all users of more-precise digital clocks have the awareness, commitment or adaptability needed to keep continually in step with Coordinated Universal Time.

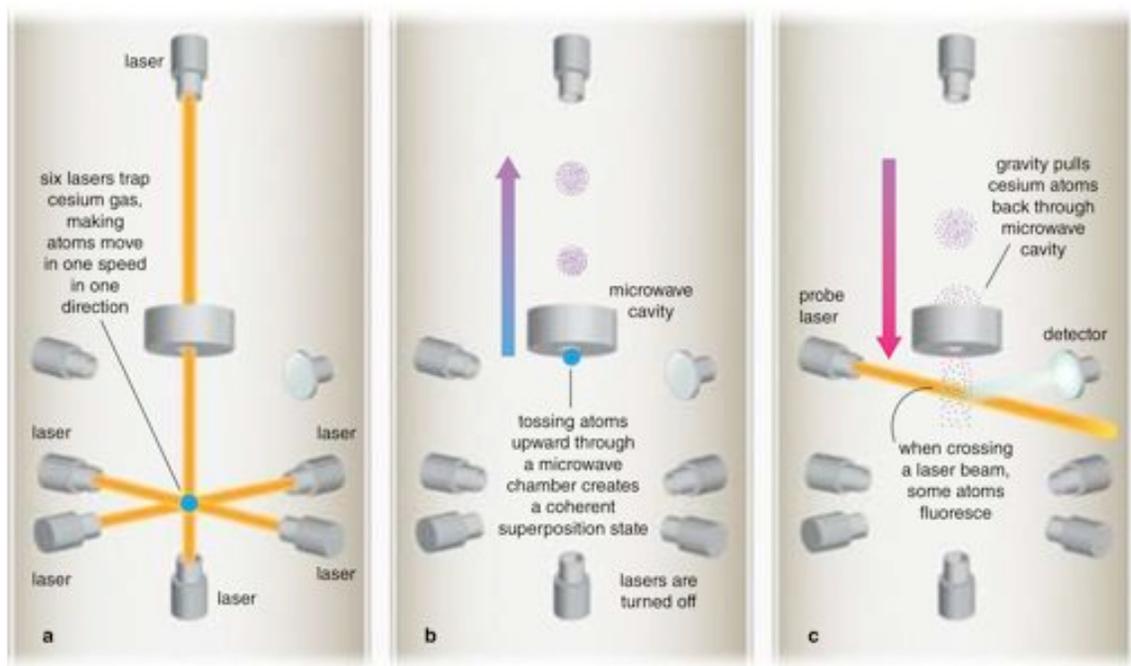

Figure 7. NIST-F1, described above, is the ultra-precise frequency standard that the National Institute of Standards and Technology in the United States uses to track atomic seconds today. It uses a fountainlike motion to manipulate cesium atoms. First, cesium gas is introduced into a vacuum chamber. Six infrared laser beams slow down the cesium atoms' movements while pushing them together into a ball. Two vertical lasers toss the ball upward about a meter high, through a microwave-filled cavity, a trip that alters the atomic states of some of the atoms. After the ball falls back down, a laser probe shines on the atoms. Those that were altered by the microwaves release light. Eventually, a microwave frequency is found that maximizes their fluorescence. That frequency is the natural resonance frequency for the cesium atom—the characteristic that defines the second. (Diagrams based on information provided by NIST.)

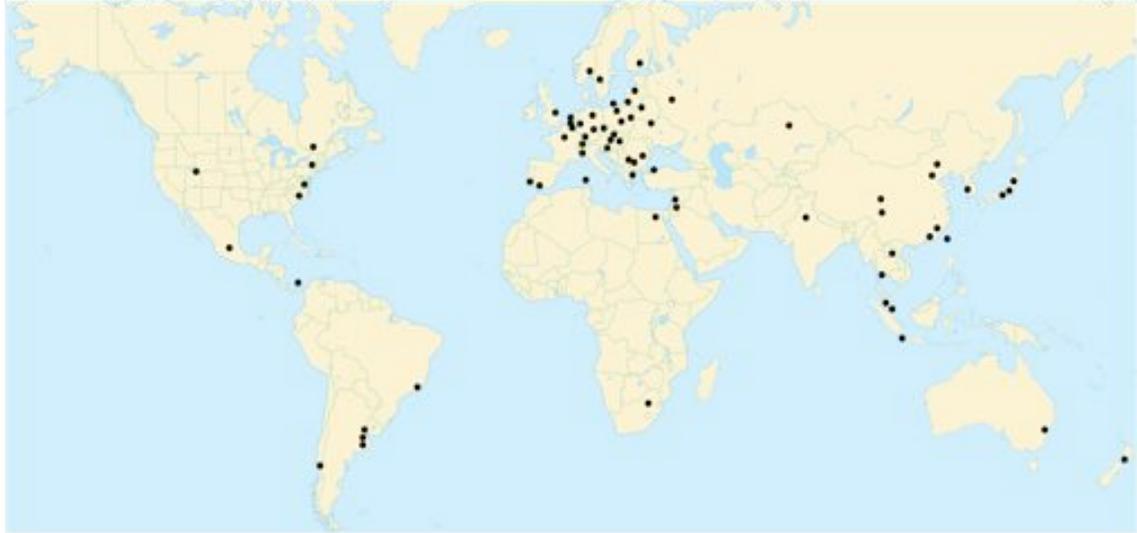

Figure 8. The atomic component of UTC is *Temps Atomique International* (TAI), which is based on atomic resonance measurements from some 200 contributors at locations, as shown above, around the world. The International Bureau of Weights and Measures (BIPM), located near Paris, maintains TAI with data received from these sites. Leap seconds are periodically introduced to adjust atomic time to keep it to within 0.9 seconds of Universal Time, adjusted for polar motion. The timing of their introduction is recommended by the International Earth Rotation and Reference Systems Service. The International Telecommunication Union, which defines the UTC broadcast recommendations, in 2012 will consider eliminating leap seconds from its definition. (Measurement device locations from BIPM website: http://www.bipm.org/en/scientific/tai/tai.html)